\documentclass[10pt,conference,letterpaper]{IEEEtran}
\IEEEoverridecommandlockouts %
\usepackage{amsmath,amssymb,graphicx,booktabs,array,microtype,cite,balance,etoolbox}
\usepackage[hidelinks]{hyperref}
\newcommand{\norm}[1]{\lVert#1\rVert}
\newcounter{calibrationalgorithm}
\newcommand{\LowShots}{262,144}
\newcommand{\HighShots}{4,194,304}
\newcommand{\LowHeld}{18.08}
\newcommand{\HighHeld}{10.06}
\newcommand{\MatrixHeld}{8.67}
\newcommand{\LowCoeff}{0.252}
\newcommand{\HighCoeff}{0.186}
\newcommand{\CircuitError}{$1.75\times10^{-17}$}

\newcommand{\SmileExact}{0.27}
\newcommand{\SmileLchs}{0.27}
\newcommand{\SkewExact}{0.26}
\newcommand{\SkewLchs}{0.26}

\newcommand{\CoarseMax}{3.55}
\newcommand{\RefinedMax}{0.036}
\newcommand{\FDMax}{0.097}

\newcommand{\lchsnode}{\eta}
\newcommand{\lchsshift}{\theta}
\makeatletter
\@ifundefined{pdfpagewidth}{}{\pdfpagewidth=8.5in \pdfpageheight=11in}
\@ifundefined{pagewidth}{}{\pagewidth=8.5in \pageheight=11in}
\makeatother
\usepackage{eso-pic}
\begin{document}
\AddToShipoutPictureFG*{\AtPageLowerLeft{\makebox[\paperwidth]{\raisebox{0.4in}{\parbox{\textwidth}{\footnotesize \copyright~2027 IEEE. Personal use of this material is permitted. Permission from IEEE must be obtained for all other uses, in any current or future media, including reprinting/republishing this material for advertising or promotional purposes, creating new collective works, for resale or redistribution to servers or lists, or reuse of any copyrighted component of this work in other works.}}}}}
\title{Quantum Calibration of Local-Volatility Surfaces from Sparse Option Quotes}
\author{\IEEEauthorblockN{Kentaro Hoshisashi}
\IEEEauthorblockA{\textit{Department of Computer Science, University College London}, London, United Kingdom\\
k.hoshisashi@ucl.ac.uk}}
\maketitle
\begin{abstract}
Calibrating a volatility surface from sparse option prices requires repeated solutions of a pricing equation as the volatility parameters change. We implement a hybrid quantum and classical method for this inverse problem using the Dupire local-volatility model. A positive coefficient representation specifies the local variance, and linear combinations of Hamiltonian simulations approximate the associated pricing evolution. The circuit propagates through successive maturity intervals and evaluates discrepancies at the observed strikes. Measurement counts provide the objective values used by a classical optimiser to update the volatility coefficients. We implement this calibration cycle in a logical gate simulator and assess it alongside classical pricing solvers and alternative surface models. Multiple-maturity synthetic examples illustrate the reconstruction of smiles and skews, while a separate small circuit experiment tests the complete measured calibration. The comparisons quantify the effects of model flexibility, pricing approximation and measurement precision. Recorded gate counts and sampling budgets identify the main computational costs of the implementation.
\end{abstract}

\section{Introduction}
The implied volatility (IV) surface represents European option prices across strikes and maturities. Observed quotes cover only part of this surface, whereas valuation and risk management require prices at unquoted contracts. Calibration therefore combines statistical fitting with financial structure. The model must reproduce observations and interpolate consistently.

The Dupire local-volatility model~\cite{dupire1994pricing} addresses this task by specifying a diffusion coefficient that depends on time and the underlying price. The associated forward equation produces option prices over the strike and maturity domain. Its calibration is an inverse problem: the coefficient field is inferred from observed prices. Implied volatility then provides a common convention for displaying those prices. Pricing under a given coefficient field is therefore distinct from calibration.

Classical methods include arbitrage-free option-price interpolation~\cite{andreasen2010volatility}, arbitrage-free smoothing~\cite{fengler2009arbitrage} and stochastic volatility inspired (SVI) parameterisation~\cite{gatheral2014arbitrage}. Physics-informed neural networks (PINNs) combine data fitting with differential equations~\cite{raissi2019physics}. Their application to volatility surfaces also incorporates financial constraints~\cite{hoshisashi2024whack}. These conditions remain relevant when a quantum solver supplies the prices.

Quantum methods have been developed for option valuation and risk estimation~\cite{stamatopoulos2020option,woerner2019quantum}. A pricing grid can also be encoded in quantum amplitudes and evolved under a discretised pricing equation. Local-volatility pricing has been implemented as quantum circuits, with corresponding estimates of their resource requirements~\cite{kaneko2022quantum}. Recent quantum methods for partial differential equations (PDEs) include local- and stochastic-volatility pricing and the reconstruction of an implied-volatility smile from quantum prices~\cite{guseynov2025quantum,guseynov2026end}. 

For calibration, the pricing operation must be connected to a criterion that compares its output with observed quotes. An explicit formulation of PDE-constrained optimisation provides this connection, with a financial example estimating one constant volatility~\cite{sato2025explicit}. The construction uses linear combinations of Hamiltonian simulations (LCHS)~\cite{an2023linear,an2026quantum} and an explicit objective encoding. Spatially varying coefficients can also be parameterised through diagonal block encoding~\cite{yano2026quantum}. Other quantum-finance pipelines calibrate a distribution classically before applying quantum integration~\cite{hok2026quantum}, while a quantum McKean--Vlasov method identifies stochastic-local-volatility calibration as a further extension~\cite{miyamoto2026quantum}.

We specialise these ideas to a local-volatility surface fitted from sparse prices at multiple maturities. Each candidate field changes the pricing operator. The circuit must retain the dependence of later prices on earlier coefficient intervals, map the price grid to the actual quoted strikes, and return a financially interpretable discrepancy. We implement the resulting objective evaluation within a classical optimisation loop, in place of the coherent optimiser of \cite{sato2025explicit}. The output is a classical vector of model coefficients.

A calibration comparison requires more than a pricing error. Different coefficient families can fit the same quotes yet imply different smiles between them. Even within a fixed family, propagation error can change butterfly prices. The experiments examine these effects separately from the variability introduced by quantum measurements.

This paper makes three contributions:
\begin{itemize}
\item We formulate multi-expiry Dupire calibration as an explicit quantum-circuit problem, including positive local variance, off-grid observations and the propagation between maturity intervals. To our knowledge, this specific inverse-problem implementation has not previously been reported.
\item We implement the full hybrid calibration cycle using finite-shot circuit measurements for coefficient updates. Independent matrix calculations verify the residual encoding, and the decomposed circuits and measurement records are provided.
\item We compare the financial output and computational requirements with classical methods. The experiments distinguish the attainable surface shapes, the accuracy of LCHS pricing within a fixed model, and the precision of the measured calibration objective.
\end{itemize}

\section{Volatility Surface Calibration}
\subsection{Calibration Problem}
Let $C(T,K)$ denote a European call price, $D(T)$ the discount factor and $F(T)$ the forward. With deterministic interest rates and proportional dividends, define the normalised strike and price by
\begin{equation}
 k=K/F(T),\qquad c(T,k)=C(T,K)/(D(T)F(T)).
 \label{eq:normalisation}
\end{equation}
\begin{figure*}[t]
\centering\includegraphics[width=\textwidth]{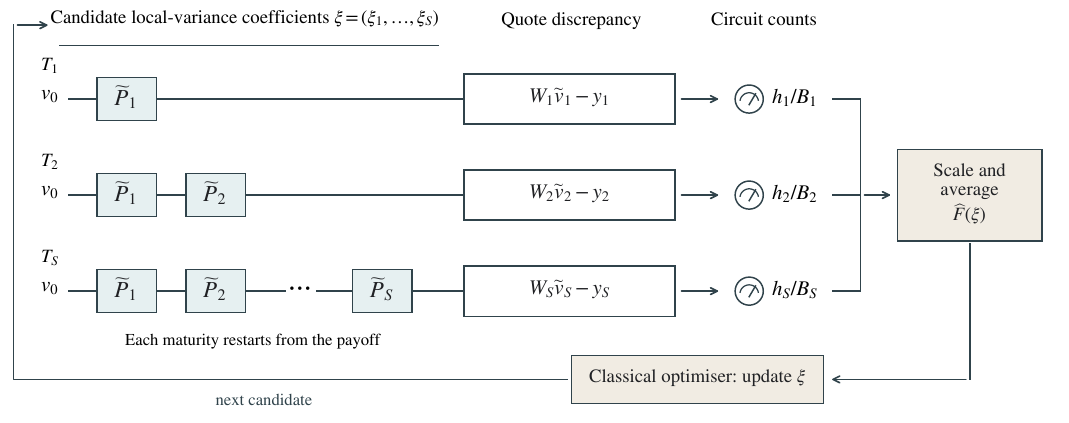}
\caption{Multi-expiry calibration. Separate maturity circuits retain their pricing histories. Scaled residual counts supply the classical coefficient update.}
\label{fig:framework}
\end{figure*}
The at-the-money forward strike is $k=1$. The local variance $a(T,k)=\sigma_{\mathrm{loc}}^2(T,k)$ determines prices through
\begin{equation}
 \partial_Tc=\tfrac12a(T,k)k^2\partial_{kk}c,\qquad c(0,k)=(1-k)^+.
 \label{eq:dupire}
\end{equation}
Calibration estimates $a$ from observed calls. Inversion of the Black formula~\cite{black1976pricing} at the resulting prices gives the implied-volatility surface.

Time is divided at the quoted maturities, $0=T_0<T_1<\cdots<T_S$. On interval $s$, we use
\begin{equation}
 a_s(k;\xi_s)=a_{\min}+\sum_{j=1}^{M}\xi_{sj}\phi_j(\log k),
 \qquad 0\leq\xi_{sj}\leq1,
 \label{eq:variance}
\end{equation}
where $\phi_j\geq0$ and $a_{\min}=0.01$. Positivity is imposed by the representation. The basis functions control different parts of the local-variance curve, and the coefficient vector can change between maturity intervals. The experiments use positive quadratic functions for the illustrative surfaces, four overlapping cosine functions in the comparative benchmark, and two functions in the small circuit implementation.

We discretise log-strike on $N=2^n$ interior nodes over $[-1.2,1.2]$. At the endpoints, the fixed call values are $1-k_{\min}$ and zero. These are finite-domain boundary approximations. Let $G(k)$ be their strike-linear interpolant and write $c=G+v$. The interior dynamics are
\begin{equation}
 \dot v=-A_s(\xi_s)v,\qquad A_s(\xi_s)=A_0+\sum_j\xi_{sj}A_j.
 \label{eq:generator}
\end{equation}
The strike derivative is discretised by a fitted difference operator, whose weights are chosen so that it annihilates constants and strike-linear functions exactly. Consequently, the boundary lift introduces no forcing term in~\eqref{eq:generator}. The initial vector $v_0$ is obtained from the sampled call payoff and the same lift.

Observed strikes need not coincide with grid nodes. We use linear interpolation in strike, including the boundary values, to define the full observation map $W_s^{\mathrm{full}}$. Its interior part is $W_s$, and the adjusted observations are $y_s=c_s^{\mathrm{obs}}-W_s^{\mathrm{full}}G$. The joint calibration criterion is
\begin{equation}
 F(\xi)=\frac1S\sum_{s=1}^{S}\frac1{q_s}
 \norm{W_sv_s(\xi_1,\ldots,\xi_s)-y_s}_2^2,
 \label{eq:objective}
\end{equation}
where $q_s$ is the number of quotes at maturity $s$. Every later price depends on the preceding coefficient intervals. This dependence is retained in the joint circuit calibration. Sequential maturity calibration, used in the larger financial benchmark, instead fits one interval at a time while propagating its fitted state onwards.

\subsection{Financial Constraints}
A normalised call surface should be decreasing and convex in strike, lie between intrinsic value and one, and increase with maturity at fixed normalised strike. In differential form, the relevant conditions are
\begin{equation}
 -1\leq\partial_kc\leq0,\qquad \partial_{kk}c\geq0,
 \qquad \partial_Tc\geq0.
 \label{eq:constraints}
\end{equation}
These conditions constrain digital, butterfly and calendar positions. Strike convexity corresponds to a nonnegative state-price density~\cite{breeden1978prices}. The maturity condition concerns the forward-normalised price in~\eqref{eq:normalisation}. It does not apply directly to nominal calls under arbitrary carry.

For positive local variance, the fitted spatial operator preserves the corresponding discrete inequalities under exact propagation and backward Euler. Linear strike interpolation preserves the strike conditions between nodes. The technical supplement gives the derivation, including the boundary assumptions. A finite LCHS approximation introduces an additional pricing error, so its output is checked numerically against the same inequalities. If the maximum price perturbation is $\delta_s$, a normalised three-strike butterfly can deviate from its exact value by at most $2\delta_s$, while a calendar difference changes by at most $\delta_s+\delta_{s-1}$. A small root mean square error (RMSE) in prices does not establish financial consistency.

\section{Quantum Calibration}
\subsection{Quantum Pricing Operator}
Figure~\ref{fig:framework} shows the maturity-indexed construction. Each residual circuit starts from the payoff and propagates through the intervals preceding its observed maturity. The resulting state enters the quote residual circuit.

Quantum gates implement unitary evolution, whereas the pricing equation is diffusive. LCHS expresses the required propagation as a weighted sum of unitary evolutions~\cite{an2023linear,an2026quantum}. Choose a coefficient-independent shift $\lchsshift$ such that the Hermitian part of $A_s+\lchsshift I$ is nonnegative, and write $A_s+\lchsshift I=L_s+iH_s$. Over an interval of length $\Delta_s$,
\begin{equation}
 \widetilde P_s=e^{\lchsshift\Delta_s}
 \sum_{\ell=1}^{J}w_\ell
 e^{-i\Delta_s(H_s+\lchsnode_\ell L_s)}.
 \label{eq:lchs}
\end{equation}
The nodes $\lchsnode_\ell$ and weights $w_\ell$ come from a finite quadrature. We use the improved kernel~\cite{an2026quantum} with $\beta=0.8$, midpoint nodes on $[-K_c,K_c]$ and weights normalised to sum to one. Table~\ref{tab:finance} calls it ``beta'', and its ``simple'' kernel is the original one of~\cite{an2023linear}. The term count $J$ and cutoff $K_c$ determine the approximation to the chosen pricing operator. No bound relating $(J,K_c)$ to the price error is derived here.

Let $\alpha=\sum_\ell|w_\ell|$. An auxiliary register is prepared with amplitudes $\sqrt{|w_\ell|/\alpha}$. A controlled operation selects the corresponding Hamiltonian evolution, including the phase of $w_\ell$, and the preparation is reversed. The all-zero auxiliary branch contains the sum in~\eqref{eq:lchs}, divided by its known normalisation. The selected evolutions are combined coherently within the circuit.

For maturity $s$, the pricing state must represent
\begin{equation}
 \widetilde v_s=\widetilde P_s\widetilde P_{s-1}\cdots\widetilde P_1v_0.
 \label{eq:propagation}
\end{equation}
We allocate a separate auxiliary register to each interval. An unsuccessful branch at one interval then remains distinguishable at all subsequent intervals. Projecting all interval registers onto zero yields the product in~\eqref{eq:propagation}. No intermediate price vector is measured or supplied by a classical solver.

\subsection{Residual Measurement}
To compare model prices with quotes, $W_s/\gamma_s$, with $\gamma_s\geq\norm{W_s}_2$, is embedded in a unitary observation map. This construction is a block encoding: a scaled matrix forms a block of a larger unitary~\cite{gilyen2019quantum}. A second controlled preparation introduces the adjusted observations $y_s$. Their interference produces a successful branch proportional to $W_s\widetilde v_s-y_s$. Its probability is
\begin{align}
 p_s(\xi)&=\frac{\norm{W_s\widetilde v_s-y_s}_2^2}{\zeta_s^2},
 \label{eq:probability}\\
 \zeta_s&=\norm{v_0}_2e^{\lchsshift T_s}\alpha^s\gamma_s+\norm{y_s}_2.
 \label{eq:scale}
\end{align}
The known scale $\zeta_s$ is independent of the candidate coefficients. It accounts for the initial state, each LCHS combination, the observation map and the reference preparation. Retaining this scale converts a measurement probability into a pricing discrepancy in the units of~\eqref{eq:objective}.

If $h_s$ successful outcomes are observed in $B_s$ repetitions, the objective estimate is
\begin{equation}
 \widehat F(\xi)=\frac1S\sum_s\frac{\zeta_s^2}{q_s}\frac{h_s}{B_s}.
 \label{eq:measured}
\end{equation}
For a fixed coefficient vector, this estimator is unbiased for the finite circuit's quadratic residual. With independent repetitions,
\begin{equation}
 \operatorname{Var}(\widehat F)=\frac1{S^2}\sum_s
 \frac{\zeta_s^4}{q_s^2B_s}p_s(1-p_s).
 \label{eq:sampling}
\end{equation}
This uncertainty is separate from the difference between finite LCHS and exact PDE propagation. Zero observed successes therefore do not confirm zero pricing error.

\subsection{Hybrid Optimisation}
Algorithm~\ref{alg:calibration} summarises the bounded pattern search used to update the coefficients from measured objectives. Starting from $\xi_{sj}=0.5$, it compares the incumbent with positive and negative steps along each coefficient direction and along opposing within-interval directions. The latter adjust the relative weight of the two local-variance basis functions in the circuit example. Candidates are clipped to $[0,1]$, duplicates are removed, and ties retain the incumbent. Two sweeps use a step of 0.25.

Every comparison remeasures the incumbent with the same repetition budget as the alternative candidates. Only measurement counts enter the coefficient decisions. Exact residuals, statevector amplitudes and the generating coefficients are used separately for validation. After selection, fresh counts provide the final residual estimate and its 95\% confidence interval (CI), independently of the measurements used to select the coefficients.

The implementation uses Qiskit and Aer~\cite{javadi2024quantum}. For these small circuits, explicit Hamiltonian unitaries and observation dilations are decomposed into single-qubit and controlled-NOT (CNOT) gates. The notebook performs circuit synthesis, sampling and optimisation; reported resources include dense synthesis. Qubitization is an alternative for Hamiltonian simulation under suitable operator access~\cite{low2019hamiltonian}, but is not used here. Logical gates assume all-to-all connectivity. Physical execution requires device-specific routing and sufficient fidelity.

\par\smallskip
\noindent\begin{minipage}{\columnwidth}
\small
\hrule\smallskip
\refstepcounter{calibrationalgorithm}\label{alg:calibration}
\textbf{Algorithm \thecalibrationalgorithm. Measured local-volatility calibration}
\par\smallskip\hrule\smallskip
\textbf{Input:} quotes and simulation budgets.
\par\smallskip
\begin{tabular}{@{}r@{\hspace{0.7em}}p{\dimexpr\columnwidth-1.8em\relax}@{}}
1 & $\xi\gets0.5$\\
2 & \textbf{for} sweep $=1,2$ and direction $d$:\\
3 & \hspace*{0.8em}$\mathcal X\gets\{\xi,\operatorname{clip}_{[0,1]}(\xi\pm0.25d)\}$\\
4 & \hspace*{0.8em}\textbf{for} $x\in\mathcal X$:\\
5 & \hspace*{1.6em}$U_s\gets\textsc{ResidualCircuit}(x_{1:s})\quad\forall s$\\
6 & \hspace*{1.6em}$h_s\gets\textsc{Measure}(U_s,B_s)\quad\forall s$\\
7 & \hspace*{1.6em}$\widehat F(x)\gets\frac1S\sum_s\frac{\zeta_s^2h_s}{q_sB_s}$\\
8 & \hspace*{0.8em}$\xi\gets\arg\min_{x\in\mathcal X}\widehat F(x)$\\
9 & $\widehat F_{\mathrm{final}},\mathrm{CI}\gets\textsc{FreshMeasurement}(\xi)$\\
10 & \textbf{return} $\xi,\widehat F_{\mathrm{final}},\mathrm{CI}$
\end{tabular}
\par\smallskip\hrule
\end{minipage}
\par\smallskip

\section{Experimental Design}
\subsection{Surface Reconstruction}
We introduce two synthetic examples to show the financial objects produced by the framework: a smile and a negatively sloped skew. Both use the positive affine representation in~\eqref{eq:variance}, with three basis functions of $x=\log k$
\begin{equation}
 \begin{aligned}
 \phi_1(x)&=0.08,\quad \phi_2(x)=0.1(x+0.45)^2,\\
 \phi_3(x)&=0.1(x-0.45)^2.
 \end{aligned}
 \label{eq:basis}
\end{equation}
The first coefficient controls the variance level. The two squared terms control the relative weight of the two wings and curvature. Equal weights on these terms produce the smile example. A larger coefficient on $(x-0.45)^2$ raises volatility at lower strikes. The generating values are specified in the notebook.

Each surface has four quoted maturities, $T=(0.25,0.5,0.75,1)$ years, and nine irregular strikes between $k=0.7$ and $1.4$. Reference prices use 1,024 interior nodes. Bounded least squares fits the three coefficients sequentially on a 128-node grid using exact matrix propagation. Exact, backward-Euler and beta-LCHS prices are then evaluated at the same fitted coefficients, with $J=8192$ and $K_c=96$ for LCHS. This common-coefficient comparison isolates forward approximation from optimisation. It is distinct from the independently calibrated solver comparison below.

\subsection{Classical Comparators}
The larger benchmark contains independent calibrations using matrix exponentiation, sparse backward Euler and two LCHS kernels within the same four-basis model. Andreasen--Huge (AH) interpolation and fixed/adaptive physics-informed neural networks provide alternative representations~\cite{andreasen2010volatility,hoshisashi2024whack}. The adaptive neural comparator is a documented implementation adaptation. All PDE solvers share the coefficient family, quotes, parameter bounds and calibration objective.

The benchmark uses $N=32,64,128$, two maturities, five optimiser starts, and fine-grid targets with 2,048 interior nodes. Baseline targets use the calibration model to generate prices. Alternative targets use the local variance $0.01+0.055\,e^{-((x+0.1)/0.3)^2}+0.035/(1+e^{4x})$ on both maturity intervals. Clean data and uniform price perturbations bounded by 5 basis points (bp) are considered, with three noise datasets for the latter. We first average optimiser starts within each dataset, then aggregate datasets. Of 568 planned runs, 567 complete. One AH input route fails because a perturbed call lies below intrinsic value and has no Black implied volatility. The noisy baseline AH result therefore averages two datasets, while all other noisy results average three. Paired comparisons are reported in the supplement.

\subsection{Circuit Simulation and Evaluation}
The measured calibration uses four interior grid nodes, two coefficients per interval, four quotes per maturity and $T=(0.5,1)$ years. Its four generating coefficients are $(0.75,0.25)$ and $(0.25,0.75)$. Targets use exact propagation on the same grid. This small instance tests the circuit construction and optimisation interface. Its coarse-grid IV shape is not used to assess the financial quality of a reconstructed market surface.

The circuit configuration has $J=8$ and $K_c=12$. Two repetition budgets, \LowShots{} and \HighShots{} per maturity per candidate, are tested with sampler seeds 101, 202 and 303. Exact, backward-Euler and finite-LCHS matrix objectives use the same pattern search. Separate statevector checks compare circuit probabilities with independently computed residuals for two coefficient vectors and both maturities. A further resource sweep constructs 12 circuits varying the grid, LCHS terms and number of intervals.

Price errors are reported in basis points of normalised call value, where 1 bp is $10^{-4}$ in $c$. Volatilities are annualised percentages. Held-out RMSE uses 141, 181 and 81 equally spaced strikes on $[0.7,1.4]$ per maturity (surfaces, benchmark, circuit), less any calibration strike, against the noise-free target prices. Model flexibility is controlled by the basis, spatial error by the pricing grid, quadrature error by $(J,K_c)$ and measurement uncertainty by $B_s$. Increasing one computational budget does not automatically reduce the other error sources. Runtime measurements distinguish classical calibration, circuit synthesis and gate simulation.

\section{Results}
\subsection{Calibrated Surfaces}
\begin{figure}[t]
\centering\includegraphics[width=\columnwidth]{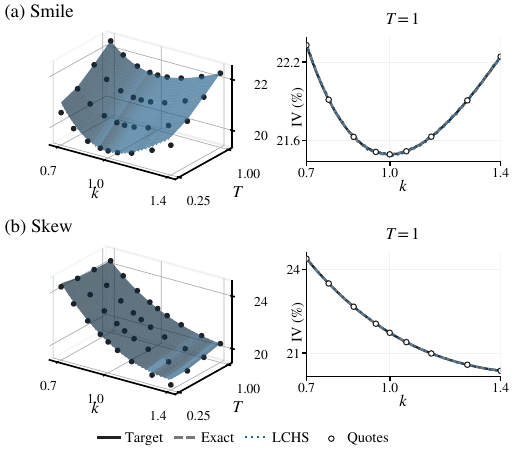}
\caption{Synthetic smile and skew: LCHS surfaces with quotes (left) and one-year comparisons (right). $k=K/F$, $T$ in years and implied volatility (IV) in percent.}
\label{fig:surfaces}
\end{figure}

Figure~\ref{fig:surfaces} shows the four-maturity examples and their one-year cross-sections. The smile has higher implied volatility in the wings, whereas the skew has higher implied volatility at lower strikes. Both shapes are produced by the positive quadratic family through different coefficient values.

At the common fitted coefficients, exact and LCHS held-out errors are \SmileExact{} and \SmileLchs{} bp for the smile, and \SkewExact{} and \SkewLchs{} bp for the skew. The curves show the effect of replacing the forward propagator while preserving the fitted model. Doubling the reference grid from 1,024 to 2,048 nodes changes quote prices by less than 0.01 bp in RMSE in each example. The close agreement is obtained with the stated 8,192-term quadrature. This agreement should not be extrapolated to the eight-term circuit or interpreted as a saving in computational cost.

\subsection{Solver Comparison}
Table~\ref{tab:finance} reports the independent calibrations. Beta LCHS stays within $0.03$ bp of exact propagation and finite differences (FD) within the same model. For the noisy alternative case, the corresponding errors are 6.60, 6.58 and 6.60 bp. The simple LCHS kernel is less accurate at the prescribed budget. Thus the choice of finite approximation matters even when the coefficient representation is unchanged. At $N=128$ (supplement) beta LCHS violates in-band butterfly convexity (Sec.~\ref{sec:consistency}).

AH and the adaptive PINN reach 3.48 and 2.06 bp on that noisy target. Their flexibility matters when the four-basis family cannot represent the target. The ranking changes for the noiseless baseline case, showing the importance of separating solver accuracy from model flexibility.

\begin{table}[t]
\caption{Price RMSE on held-out synthetic quotes (bp), $N=64$. Column minima in bold.}
\label{tab:finance}\centering
\footnotesize\setlength{\tabcolsep}{3.6pt}
\begin{tabular}{lrrrr}\toprule
 & \multicolumn{2}{c}{Baseline case} & \multicolumn{2}{c}{Alternative case}\\
Method & Noiseless & Noisy & Noiseless & Noisy\\\midrule
Matrix exponential & 1.08 & 1.91 & 6.42 & 6.58\\
Finite differences & 1.08 & 1.92 & 6.44 & 6.60\\
LCHS beta & 1.07 & 1.92 & 6.45 & 6.60\\
LCHS simple & 20.80 & 20.80 & 9.28 & 9.33\\
Andreasen--Huge & \textbf{0.35} & 2.73 & 1.83 & 3.48\\
PINN fixed & 1.30 & \textbf{1.73} & 1.17 & 2.12\\
PINN adaptive & 1.35 & 1.78 & \textbf{1.14} & \textbf{2.06}\\\bottomrule
\end{tabular}
\end{table}

\subsection{Measured Calibration}
\begin{table}[t]
\caption{Small-circuit calibration. Held-out price RMSE is in bp. Measured rows average three seeds; n/a means not applicable.}
\label{tab:measured}\centering
\footnotesize\setlength{\tabcolsep}{4pt}
\begin{tabular}{lrrr}\toprule
Objective & Shots/maturity & Price RMSE & Coefficient RMSE\\\midrule
Exact & n/a & 0.00 & 0.000\\
Finite differences & n/a & 0.03 & 0.000\\
Matrix LCHS & n/a & 8.67 & 0.125\\
Measured LCHS & $2^{18}$ & 18.08 & 0.252\\
Measured LCHS & $2^{22}$ & 10.06 & 0.186\\\bottomrule

\end{tabular}
\end{table}

The circuit residual probabilities agree with independent matrix calculations to a maximum absolute error of \CircuitError{}. This verifies the implemented observation and multi-expiry propagation at the tested coefficient vectors.

The exact-objective pattern search recovers the generating coefficients in the small controlled example, which lie on its lattice (start $0.5$, step $0.25$). The finite-LCHS matrix objective gives a held-out RMSE of \MatrixHeld{} bp. The measured searches give mean held-out errors of \LowHeld{} and \HighHeld{} bp at the two repetition budgets, respectively. These averages include all three declared seeds. More repetitions lower the mean here, though three seeds establish no trend and finite sampling also changes the sequence of selected candidates.

\begin{figure}[!t]
\centering\includegraphics[width=\columnwidth]{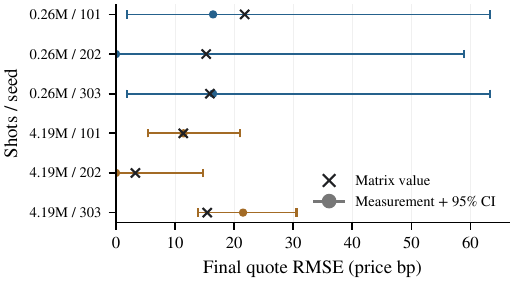}
\caption{Independent residual measurements with 95\% confidence intervals (CIs); M denotes million shots. Crosses show finite-LCHS values at the same coefficients.}
\label{fig:precision}
\end{figure}
Figure~\ref{fig:precision} evaluates the precision of the final residual measurement. The error bars are obtained from independent final observations, using conservative binomial intervals and the normalisation in~\eqref{eq:scale}. The intervals are expressed as root mean squared price errors for financial interpretation. They are appreciably wider at the lower repetition budget. A zero observed residual can coexist with a positive interval upper limit, so it does not establish an exact fit.

Price fit and coefficient identification need separate assessment. Mean coefficient RMSE falls from \LowCoeff{} to \HighCoeff{} at the higher repetition budget (Table~\ref{tab:measured}). However, the seed with the lowest held-out price error, 2.99 bp, has coefficient RMSE 0.217, above the matrix search's 0.125. Better price fit need not recover the generating coefficients more accurately. The measurement intervals concern residuals at fixed coefficients, not parameter identification.

\subsection{Pricing Accuracy and Financial Consistency}
\label{sec:consistency}
The financial checks remain part of this evaluation. All 120 exact and 120 backward-Euler outputs satisfy the counted butterfly and calendar inequalities. All 120 raw beta-LCHS outputs fail at least one inequality of each type over the full truncated domain at tolerance $10^{-8}$. At $N=32,64$, violations occur outside the quoted band. At $N=128$, 15 of 40 beta runs also fail butterfly convexity inside it, with a maximum in-band violation of 3.06 bp. Exact re-evaluation at the fitted coefficients removes these violations. The small RMSE difference between solvers should therefore be interpreted together with the derivative checks.

\begin{figure}[t]
\centering\includegraphics[width=\columnwidth]{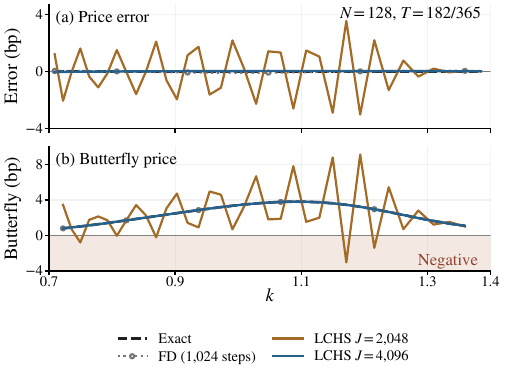}
\caption{Fixed-coefficient stress case. More LCHS terms reduce price error (top) and remove negative in-band butterflies (bottom). Shading marks negative butterfly prices.}
\label{fig:consistency}
\end{figure}

Figure~\ref{fig:consistency} examines the clean outside-model run attaining the largest quoted-band butterfly deficit ($N=128$, start 0, first expiry). All curves use its same fitted coefficients. The upper panel measures propagation error against exact prices, so model-fitting error is excluded. The lower panel prices a three-strike butterfly: a strike-weighted average of the two neighbouring calls minus the middle call. Its nonnegative payoff requires a nonnegative price.

Doubling $J$ from 2,048 to 4,096 at fixed $K_c=48$ reduces the maximum in-band price error from \CoarseMax{} to \RefinedMax{} bp, compared with \FDMax{} bp for backward Euler. The five negative in-band butterflies disappear. Violations remain outside the band, so this is a local improvement in one stress case, not a surface-wide guarantee. Increasing the term count improves price accuracy and local convexity in this example. It also increases the number of controlled evolutions in the LCHS sum.

\label{sec:resources}
\begin{table}[t]
\caption{Logical resources per residual circuit, before sampling.}
\label{tab:resources}\centering
\footnotesize\setlength{\tabcolsep}{4.4pt}
\begin{tabular}{rrr r r r}\toprule
 & & \multicolumn{2}{c}{One interval} & \multicolumn{2}{c}{Two intervals}\\
$N$ & $J$ & Qubits & CNOT & Qubits & CNOT\\\midrule
4 & 4 & 6 & 2,260 & 8 & 4,322\\
4 & 8 & 7 & 18,432 & 10 & 36,666\\
4 & 16 & 8 & 153,438 & 12 & 306,678\\
4 & 32 & 9 & 1,256,996 & 14 & 2,513,794\\
8 & 4 & 7 & 10,090 & 9 & 19,200\\
8 & 8 & 8 & 77,590 & 11 & 154,200\\\bottomrule

\end{tabular}
\end{table}

\noindent\textbf{Computational resources.} The resource requirements extend beyond the price register. A residual circuit for $s$ intervals uses
\begin{equation}
 n+s\lceil\log_2J\rceil+2
 \label{eq:qubits}
\end{equation}
logical qubits in the prototype, including distinct LCHS registers, an observation flag and a residual branch. With $N=4$ and $J=8$, one- and two-interval circuits use 7 and 10 qubits, with 18,432 and 36,666 CNOT gates. Increasing $J$ to 32 raises the two-interval circuit to 14 qubits and 2,513,794 CNOT gates under the implemented dense decomposition. Table~\ref{tab:resources} records this gate cost alongside the state and auxiliary qubits.

Separate residual circuits repeat maturity prefixes, requiring $S(S+1)/2$ interval propagations per candidate. The normalisation $\zeta_s$ contains $e^{\lchsshift T_s}\alpha^s$, which affects the repetitions needed to resolve a financial error through~\eqref{eq:sampling}. The shot-weighted gate totals include this sampling cost. They describe logical workloads, without estimating hardware runtime.

In the larger noisy outside-model benchmark, mean classical calibration times are approximately 0.0712 seconds for matrix exponentiation, 0.366 seconds for finite differences and 70.5 seconds for the beta-LCHS simulation. The latter is a classical evaluation of the quantum-compatible formula. It does not measure a quantum device's execution time.

\section{Discussion}
The residual circuit returns a scalar pricing discrepancy without reconstructing the price grid. It retains preceding maturity intervals and maps prices to the observed strikes for each proposed coefficient vector.

Amplitude estimation can improve sampling precision per oracle call under coherent access to the preparation circuit and its inverse~\cite{brassard2000quantum,montanaro2015quantum}. Our independent-shot implementation does not include this improvement. An alternative sampler would require a new depth and repetition analysis at the same price-error tolerance.

A computational benefit requires propagation and readout savings to exceed coefficient access, state preparation, synthesis and sampling costs. Sparse finite differences with adjoint derivatives provide a classical comparator (timings in Sec.~\ref{sec:resources}).

The experiments use synthetic targets and noiseless logical simulation with finite sampling. Market-data calibration, unrestricted coefficient fields, device noise, routing and fault-tolerant overhead remain to be assessed.

\section{Conclusion}
We implement local-volatility calibration from sparse quotes using quantum residual measurements and classical coefficient updates. Positive coefficient fields, off-grid observations and inter-maturity dependence are encoded explicitly. The comparisons distinguish surface flexibility, propagation accuracy and measurement precision. Refining LCHS improves local financial consistency in a fixed-coefficient stress case, while the logical circuit costs identify what must improve before a computational advantage can be established.

\section*{Acknowledgement}
Generative AI (Claude, Anthropic~\cite{anthropic2026claude}) assisted in drafting and editing all sections and in preparing the public code. The author verified all content and takes full responsibility for it.

\noindent\textbf{Data and code.}
The notebook, scripts, run logs, circuits, measurement records and the technical supplement are available at \url{https://github.com/khoshisashi/quantum-local-vol}.

\clearpage
\IEEEtriggeratref{14}
\bibliographystyle{IEEEtran}
\bibliography{references}
\end{document}